# Instability, Rupture and Fluctuations in Thin Liquid Films: Theory and Computations

**Miguel A. Durán-Olivencia**[1] · **Rishabh S. Gvalani**[2] · **Serafim Kalliadasis**[1] · **Grigorios A. Pavliotis**[2]

**Abstract**
Thin liquid films are ubiquitous in natural phenomena and technological applications. They have been extensively studied via deterministic hydrodynamic equations, but thermal fluctuations often play a crucial role that needs to be understood. An example of this is dewetting, which involves the rupture of a thin liquid film and the formation of droplets. Such a process is thermally activated and requires fluctuations to be taken into account self-consistently. In this work we present an analytical and numerical study of a stochastic thin-film equation derived from first principles. Following a brief review of the derivation, we scrutinise the behaviour of the equation in the limit of perfectly correlated noise along the wall-normal direction, as opposed to the perfectly uncorrelated limit studied by Grün et al. (J Stat Phys 122(6):1261–1291, 2006). We also present a numerical scheme based on a spectral collocation method, which is then utilised to simulate the stochastic thin-film equation. This scheme seems to be very convenient for numerical studies of the stochastic thin-film equation, since it makes it easier to select the frequency modes of the noise (following the spirit of the long-wave approximation). With our numerical scheme we explore the fluctuating dynamics of the thin film and the behaviour of its free energy in the vicinity of rupture. Finally, we study the effect of the noise intensity on the rupture time, using a large number of sample paths as compared to previous studies.

**Keywords** Thin liquid films · Dewetting · Thermal fluctuations · Fluctuating hydrodynamics

✉ Serafim Kalliadasis
s.kalliadasis@imperial.ac.uk

Miguel A. Durán-Olivencia
m.duran-olivencia@imperial.ac.uk

Rishabh S. Gvalani
rishabh.gvalani14@imperial.ac.uk

Grigorios A. Pavliotis
g.pavliotis@imperial.ac.uk

1 Department of Chemical Engineering, Imperial College London, London SW7 2AZ, UK

2 Department of Mathematics, Imperial College London, London SW7 2AZ, UK

## 1 Introduction

A thin liquid film can be understood as a layer of liquid with thickness ranging from fractions of a nanometer to several micrometers, typically resting or flowing on a substrate. Thin liquid films are quite often found in nature and nowadays technological applications: from gravity currents under water and lava flows, falling films of water on the surface of windows and sloppy streets on a rainy day, to films used in industrial coating processes for decorative, insulating or protective purposes and the cooling of microelectronic devices, to name but a few examples [10,36,44,57,61]. Thin films have attracted considerable attention since the pioneering work of Reynolds on lubrication [68], who realised their significance in both applications and fluid dynamics fundamentals. Over the last few decades, understanding the behaviour of thin films and being able to make reliable predictions of their stability and dynamics has been crucial in the rapidly growing field of microfluidics, i.e. in the art of miniaturizing chemical and lab-on-chip devices (e.g. Ref. [65]). Such small-scale setups have shown a tremendous relevance to model and replicate biological systems, e.g. blood circulation systems [8,66,72], or biological processes, such as in-vivo protein crystallisation and bone formation [2,31,47].

It is therefore no surprise that the problem of modelling and predicting the behaviour of thin liquid films has attracted the attention of numerous researchers from engineering, physics, and applied mathematics, over the last few decades. In most practical cases, the thin films are subject to additional effects and complexities, such as body forces, often gravity as in the case of falling liquid films [18,19,44], three-phase moving contact lines [56,73,74], thermo-/solutocapillary Marangoni effects [43,60] and non-Newtonian effects [63,71]. The present study focuses on the particular case of a thin film of Newtonian fluid, resting on a planar horizontal substrate and driven by the competition between surface tension and intermolecular forces (but nevertheless, the theoretical-computational framework we develop and corresponding conclusions are general, and, with proper minor modifications, are expected to be applicable to a wide spectrum of physical problems dominated by fluctuations).

The most fundamental approach to thin films begins with the general equations of hydrodynamics, usually referred to as Navier–Stokes equations of motion. These equations represent the conservation laws for the observable fields, i.e. density, velocity and energy, which are typically written as a set of partial differential equations (PDEs). At constant temperature, they reduce to two physical conservation laws: the conservation of mass and the conservation of linear momentum. The complexity of these equations can be considerably reduced by utilising the so-called long-wave approximation (or lubrication approximation for vanishing Reynolds numbers), typically valid for small slopes and strong surface tension effects. The cornerstone of the long-wave approximation is the disparity in scale between the characteristic film height $d$ along the wall-normal/cross-stream direction and the characteristic length scale $\lambda$ along the parallel-to-the-wall/streamwise direction which is taken to be long, so that the so-called film parameter, $\varepsilon = d/\lambda \ll 1$, a consequence of the strong coherence of the film in the cross-stream direction due to the action of viscosity. Rescaling the hydrodynamic problem by using the characteristic parameters $d$, $\lambda$ and $\varepsilon$, a long-wave expansion, i.e. a regular perturbation expansion for $\varepsilon$, leads to a model of reduced dimensionality, a single nonlinear PDE of the evolution type for the film thickness $h(x, t)$ where $x$ is the streamwise direction and $t$ denotes times. Such an equation is often referred to as the thin-film equation, and although it is much simpler than the full hydrodynamic problem, it still captures its basic features. Detailed descriptions of the long-wave approximation can be found in Refs. [44,54,57].

This procedure is appropriate for deterministic systems. But thin films are often susceptible to thermal fluctuations, especially as the film thickness approaches the nanoscale [41]. To include such effects into the thin-film equation we can follow the same procedure for the deterministic case but starting from a different point so that fluctuations are self-consistently taken into account. Namely, we must start by considering the equations of fluctuating hydrodynamics (FH), originally postulated by Landau and Lifshitz [48] on the basis of phenomenological arguments, and then proceed with the long-wave approximation as discussed above. But it should be noted that the phenomenological origin of FH should not be seen as a weakness, as indeed a great deal of theoretical work has been successfully undertaken to derive it from first principles since the days of Landau and Lifshitz (an extensive review is given in Ref. [32]). After applying the long-wave approximation to FH, one obtains a stochastic PDE (SPDE) for the film height which we will refer to as the stochastic thin-film equation. This model was almost simultaneously proposed by Davidovitch et al. [12] and Grün et al. [37], using phenomenological and formal mathematical arguments, respectively.

The stochastic thin-film equation exhibits the convenient structure of a conservative stochastic gradient flow. This means that the deterministic part of the equation evolves according to the steepest descent of an energy functional $\mathcal{H}(h)$ [67,76,78] and the fluctuation term is a conservative field. As discussed by Grün et al. [37], the conservative structure makes the noise independent of the stochastic calculus chosen even though it is state dependent, with Itô's and Stratonovich's calculus being the most widely used [28,69]. This equivalence is quite important as we do not have to worry about the correct interpretation of the noise, which occurs more often than not in stochastic modelling [30,69]. However, the noise appearing in the stochastic thin-film equation involves a convolution integral along the wall-normal direction. And this appears to be against the spirit of the long-wave approximation which aims at removing the dependency of the time-evolution equation on the cross-stream coordinate. Surprisingly, the work of Davidovitch et al. [12] does not include explicitly such a complicated fluctuating term, despite starting from the same point, i.e. the FH equations. In fact, these authors postulated an SPDE with a much simpler multiplicative noise almost without theoretical justification and motivated primarily by physical arguments. Contrarily, Grün et al. [37] realised the need to alleviate the complexity of the noise which they replaced with a conservative state-dependent version as well. But instead of postulating the simpler noise, Grün et al. discretised the original SPDE to transform it into a set of stochastic differential equations (SDEs). Analysing the associated Kramers-Moyal coefficients [28,69] they found an alternative noise that converges to the same Fokker–Planck equation, hence implying the same statistics. Such an equivalent noise term can then be replaced in the original SPDE yielding the equation in Ref. [12], the most widely-accepted version of the stochastic thin-film equation (e.g. [27,53,65]). This equation opens the door to numerical scrutiny of the effect of noise on dewetting.

The present work introduces an alternative and more systematic derivation of the stochastic thin-film equation from FH. Our derivation differs from the previous works fundamentally in two points. First, we do not require the noise to be delta-correlated in both the streamwise and the wall-normal direction, unlike the work of Grün et al. Indeed, we obtain the same SPDE by only imposing that the noise is perfectly correlated along the cross-stream direction. We believe that this condition is more physically meaningful than imposing an uncorrelated noise along the cross-stream direction, as is the case in previous works. Indeed, as we already emphasised, it is precisely the coherence of the film in this direction due to the action of viscosity that forms the basis of the long-wave approximation. The correlation coefficients along the streamwise direction are ultimately obtained by imposing the detailed-balance condition. Second, we present an efficient numerical scheme based on spectral collocation methods, which we believe to be more convenient as it gives us the opportunity of selecting in a straightforward way the frequency modes of the noise (following the spirit of the long-wave approximation). Along the way, we also perform the necessary theoretical analysis that is missing in the literature and provide a formal definition of the rupture time. This includes a detailed linear stability analysis of the initially uniform thin film, which is necessary to understand the origin and nature of the thin-film rupture. We show that the film is unconditionally stable to sufficiently small perturbations in the case of a negligible interface potential. In the case of a general interface potential, $\phi(h) = \alpha h^{-3} - \beta h^{-2}$, which is the sum of a non-negative convex and a concave term, we find that the condition required for the film to become linearly unstable is $\beta > 2\alpha$. We also prove that the stochastic thin-film equation with the simpler multiplicative noise fulfils the detailed-balance condition, which is required from thermodynamical arguments [12,37], for a very particular noise structure. Moreover, we present simulations of the dynamics of rupture by using our numerical scheme and study how the noise intensity affects the rupture time. Our results and definition of rupture time seem to perfectly agree with the numerical results obtained by Grün et al. Our work thus brings closure, from the theoretical point of view but also in the practical sense, to the stochastic thin-film problem and provides a general theoretical and numerical apparatus that will be relevant in a wide variety of natural processes which involve a similar stochastic gradient-flow structure.

In the same spirit, we believe that our work is of relevance not only for the thin-film community, for obvious reasons, but also to the current state-of-the-art in stochastic modelling and statistical physics. This is due to the fact that the theoretical analysis and the numerical method introduced here can be useful to research fields involving a similar SPDE, which can be found in a wide variety of problems. Examples of these are the study of nanojets via phenomenological equations [25,55], the dynamics of colloidal systems and/or systems with long-range interactions via a fluctuating dynamic density-functional theory (DDFT) [1,6,7,16,23,45,46], phase transitions in complex systems with coarse-grained models from fluctuating DDFT [50,51], crystallisation via stochastic phase-field-crystal (PFC) models [26,33,34], and many more. Indeed, in recent years there has been an explosion of interest in the formal study of PDEs and SPDEs of the same structure, ever since the pioneering work of Otto [58], where it was shown that the solutions of such kind of equations for a particular mobility tensor represent a gradient flow of the Dirichlet free energy with respect to the 2-Wasserstein metric (which metrizes the weak topology of probability measures [29,77]) when the problem is formulated into a discrete time variational scheme. Similar results were obtained by Dolbeault et al. [22] and Lisini et al. [49] for concave mobility operators, but do not exist for the kind of mobility tensor involved in this work. More recently, Reina and Zimmer [67] found a general fluctuation-dissipation relation between the drift and the fluctuating term of a general stochastic gradient flow which also includes the stochastic thin-film equation considered here, so that both the maximum-entropy production (MEP) and the large-deviation (LD) principle are satisfied. Such a relation is indeed satisfied by the structure of the noise we obtain from imposing the detailed-balance condition to the equivalent multiplicative noise. It is therefore evident that theoretical and numerical methods presented in this work should be of interest to practitioners of many other fields, who might be interested in analysing the sensitivity to fluctuations or in simulating the dynamics of the system at hand, as already noted earlier.

In Sect. 2 we present the Landau–Lifshitz FH equations for a fluid film flowing on a horizontal substrate. We continue with the long-wave (or lubrication) approximation, for which we appropriately nondimensionalise the FH equations, and subsequently take a regular perturbation expansion for $\varepsilon \ll 1$. This yields the stochastic thin-film equation with a conservative noise involving a convolution integral. A flow diagram to summarise our derivation and the relationship with previous approaches is given in Fig. 1. Before substituting the convoluted noise with a simpler multiplicative version, the gradient flow structure of the deterministic part of the equation is shown in Sect. 3. In this section we also give the effective energy functional related to the thin-film equation and study the linear stability of an initially uniform thin film. In Sect. 4 a simpler multiplicative noise with free parameters, which is equivalent to the one derived from first principles, is proposed. We then study the implications of imposing the detailed-balance condition to the resultant SPDE and show the equivalence with the original noise in the limit of perfect correlation along the cross-stream direction. In Sect. 5 we introduce a numerical method to integrate the stochastic thin-film equation based on a spectral collocation scheme. Finally, concluding remarks and discussion are offered in Sect. 6.

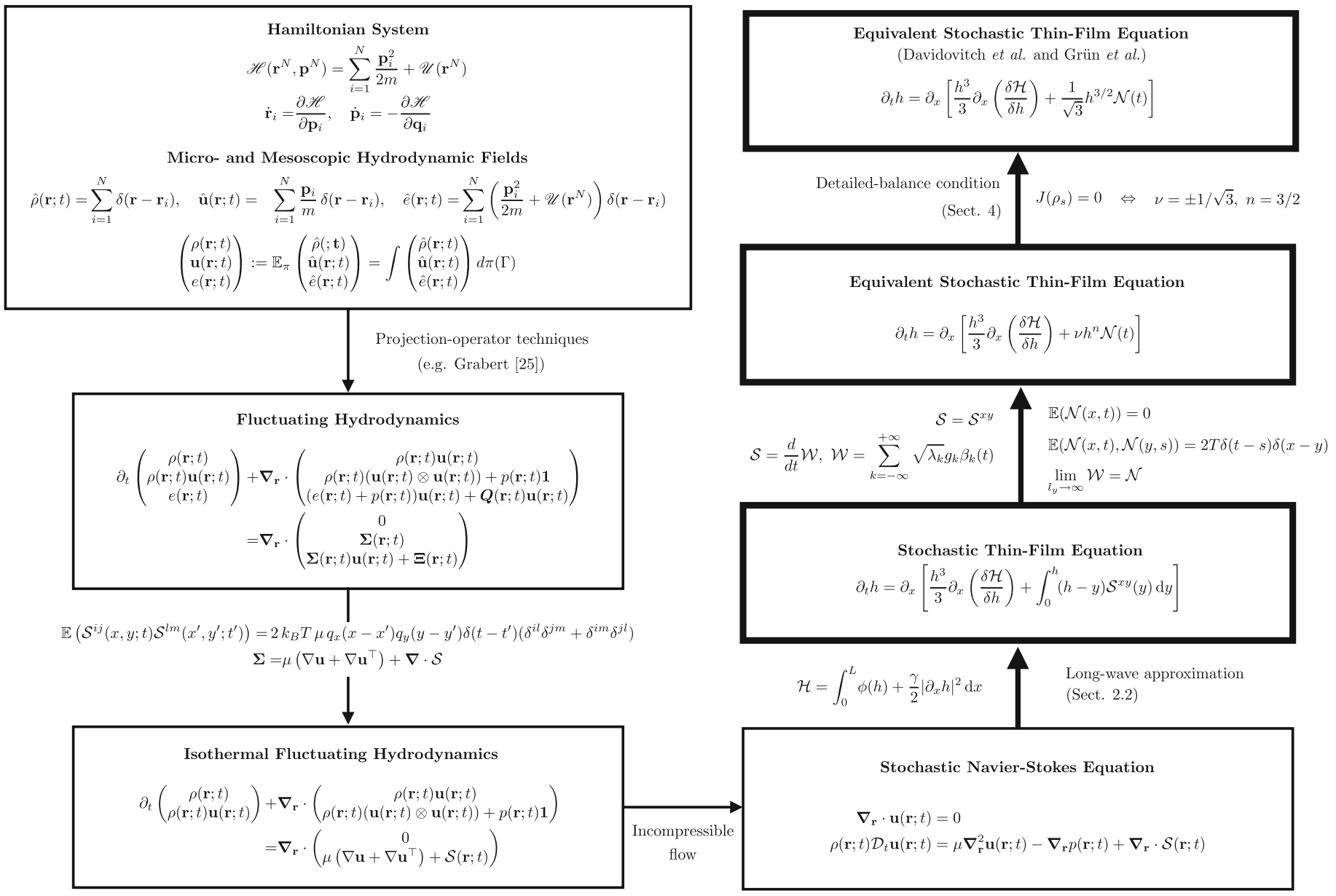

**Fig. 1** Flow diagram of the approach used to derive the stochastic thin-film equation from the full underlying Hamiltonian dynamics. Arrows indicate the interconnectedness of the different approaches. Thick black boxes/arrows: this work. Thin boxes/arrows: previous works. Text on arrows gives brief descriptions of the approximations/manipulations made

# 2 Theoretical Framework

## 2.1 Stochastic Navier–Stokes Equation

In this work, we restrict ourselves to two dimensions as in Ref. [37]. The fluctuating dynamics of a two-dimensional thin film of Newtonian fluid flowing on a horizontal substrate can be described by the incompressible Navier–Stokes equations [37] (see also Fig. 1):

$$\nabla \cdot \mathbf{u}(\mathbf{r};t) = 0, \tag{1a}$$

$$\rho(\mathbf{r};t)\,\mathcal{D}_t\mathbf{u}(\mathbf{r};t) = \mu\nabla^2\mathbf{u}(\mathbf{r};t) - \nabla p(\mathbf{r};t) + \nabla\cdot\boldsymbol{\mathcal{S}}(\mathbf{r};t), \tag{1b}$$

where $\mathbf{u} = (u, v)^\top$ is a two-dimensional velocity vector field, with $u$ and $v$ being the streamwise and cross-stream components, respectively, $\mu$ is the dynamic viscosity, $\rho$ is the fluid density, and $p$ is the pressure of the fluid. Although one would expect significant differences in the analysis, numerics and physics between two- and three-dimensional films, we believe that the simpler two-dimensional case is a good starting point, offering useful information-insight and the theoretical framework for the substantially more involved three-dimensional problem. And although events of interest, such as dewetting and subsequent drop formation, actually occur in a three-dimensional setting, we expect the phenomenology of the two-dimensional case to share some common features with that of the three-dimensional one. The operator $\mathcal{D}_t = (\partial_t + \mathbf{u}\cdot\nabla) = (\partial_t + u\partial_x + v\partial_y)$ is the convective derivative, and $\boldsymbol{\mathcal{S}}$ is the fluctuating stress tensor, which represents the effect of random thermal fluctuations on the film dynamics. The stress tensor $\boldsymbol{\mathcal{S}}$ is symmetric and has zero mean, besides having the following correlation structure:

$$\mathbb{E}\left(\mathcal{S}^{ij}(x, y; t)\mathcal{S}^{lm}(x', y'; t')\right) = 2\, k_B T\, \mu\, q_x(x-x') q_y(y-y')\delta(t-t')(\delta^{il}\delta^{jm} + \delta^{im}\delta^{jl}), \tag{2}$$

where $k_B$ is the Boltzmann constant, $T$ is a positive constant fixing the temperature/noise intensity of the system and the functions $q_x$ and $q_y$ are left undefined for the time being. At the wall, we apply the standard no-slip boundary condition:

$$\mathbf{u}(\mathbf{r}; t) = 0, \quad \forall \mathbf{r} \in \{(x, y) \in \mathbb{R}^2 :\ y = 0\}, \tag{3}$$

whereas at the fluid-air interface $y = h(x; t)$, with $h$ being the film height at the position $x$ and time $t$, we apply the stress-balance boundary condition:

$$(\boldsymbol{\sigma} + \boldsymbol{\mathcal{S}}) \cdot \hat{\mathbf{n}} = (\Pi + \gamma\,\kappa)\,\hat{\mathbf{n}} \quad \text{at} \quad y = h(x; t), \tag{4}$$

where $\boldsymbol{\sigma}$ is the viscous stress tensor, $\kappa$ is the mean curvature of the surface, $\gamma$ is the surface tension coefficient, $\hat{\mathbf{n}}$ is the normal vector to the interface, and $\Pi = -\phi'(h) = -\frac{\partial\phi(h)}{\partial h}$ is the disjoining pressure, with $\phi(h)$ the interface potential, which models molecular interactions between liquid molecules and air [37]. Additionally, we apply the following kinematic boundary condition at the interface:

$$\partial_t h = v - u\,\partial_x h \quad \text{at} \quad y = h(x; t), \tag{5}$$

which states that a fluid particle on the interface will remain there for all times, thus preventing matter from leaving the interface via e.g. evaporation or any other mechanism.

## 2.2 Long-Wave Approximation: The Stochastic Thin Film Equation

In the following we simplify the FH equations by nondimensionalising them via the parameters shown in Table 1 following Ref. [37]. In this way, we rewrite the equations in terms of two fundamental parameters, the characteristic film height $d$ along the cross-stream direction and the characteristic length scale $\lambda$ along the streamwise direction. We then take the limit $\varepsilon = d/\lambda \ll 1$ and retain terms up to and including $\mathcal{O}(\varepsilon)$. This step, and as already mentioned in the Introduction, is what is widely known as the long-wave approximation. In Table 1, $U$ is the characteristic velocity scale of the flow and quantities with tildes are non-dimensional and taken to be of $\mathcal{O}(1)$ with respect to $\varepsilon$.

The scaling of the deterministic terms follows e.g. from the analysis introduced by Dussan and Davis [13]. The noise tensor is scaled in such a way that it is of the same order as the corresponding leading order terms in the viscous stress tensor. This ensures that the terms are retained to leading order in $\varepsilon$. For the Reynolds number, $\mathrm{Re} = \rho U d/\mu$, we assume $\mathrm{Re} \sim \mathcal{O}(1)$. Thus, the equations for the streamwise and cross-stream velocities are given by:

$$\varepsilon\,\mathrm{Re}\,\mathcal{D}_t u = \left(\varepsilon^2\partial_x^2 + \partial_y^2\right)u - \partial_x(p + \Pi) + \varepsilon^2\partial_x\mathcal{S}^{xx} + \partial_y\mathcal{S}^{xy}, \tag{6a}$$

$$\varepsilon^3\,\mathrm{Re}\,\mathcal{D}_t v = \varepsilon^2\left(\varepsilon^2\partial_x^2 + \partial_y^2\right)v - \partial_y(p + \Pi) + \varepsilon^2\partial_x\mathcal{S}^{xy} + \varepsilon^2\partial_y\mathcal{S}^{yy}, \tag{6b}$$

where the tildes were removed for the sake of simplicity. To leading order in $\varepsilon$, this reduces to:

$$0 = -\partial_x(p + \Pi) + \partial_y\left(\partial_y u + \mathcal{S}^{xy}\right), \tag{7a}$$

$$0 = -\partial_y(p + \Pi). \tag{7b}$$

**Table 1** Nondimensionalisation of the FH equation, as proposed by Grün et al. [37]

| | | |
|---|---|---|
| $x = \lambda\,\tilde{x}$, | $y = d\,\tilde{y}$, | $p = \left(\frac{U\mu}{d\varepsilon}\right)\tilde{p}$, |
| $\partial_x = (1/\lambda)\,\tilde{\partial}_x$ | $\partial_y = (1/d)\,\tilde{\partial}_y$ | $\Pi = \left(\frac{U\mu}{d\varepsilon}\right)\tilde{\Pi}$ |
| $u = U\,\tilde{u}$ | $v = \varepsilon\,U\tilde{v}$ | $t = \left(\frac{\lambda}{U}\right)\tilde{t}$ |
| $\gamma = \left(\frac{U\mu}{\varepsilon^3}\right)\tilde{\gamma}$ | $\kappa = \left(\frac{\varepsilon^2}{d}\right)\tilde{\kappa}$ | $h = d\,\tilde{h}$ |
| $\mathcal{S}^{xy} = \left(\frac{U\mu}{d}\right)\tilde{\mathcal{S}}^{xy}$ | $(\mathcal{S}^{xx}, \mathcal{S}^{yy}) = \left(\frac{U\mu}{\lambda}\right)(\tilde{\mathcal{S}}^{xx}, \tilde{\mathcal{S}}^{yy})$ | $T = \left(\frac{\mu U\lambda^2}{\varepsilon k_b}\right)\tilde{T}$ |

On the other hand, the continuity equation (1a), no-slip condition (3) and the kinematic boundary condition (4) remain unchanged. It can also be shown that the curvature $\kappa$ becomes $\partial_x^2 h + \mathcal{O}(\varepsilon^2)$, while the normal component of the normal stress tensor, $\hat{\mathbf{n}}\cdot(\sigma + \mathcal{S})\cdot\hat{\mathbf{n}}$, is simply $-p + \mathcal{O}(\varepsilon^2)$. Therefore, to leading order in $\varepsilon$, the interface stress balance condition (5) reduces to:

$$p = -\gamma\,\partial_x^2 h \quad \text{at} \quad y = h. \tag{8}$$

The tangential component of the normal stress, given by $\hat{\mathbf{t}}\cdot(\boldsymbol{\sigma} + \boldsymbol{\mathcal{S}})\cdot\hat{\mathbf{n}}$, then becomes:

$$\hat{\mathbf{t}}\cdot(\boldsymbol{\sigma} + \boldsymbol{\mathcal{S}})\cdot\hat{\mathbf{n}} = \frac{\left(\partial_y u + \mathcal{S}^{xy}\right)\partial_x h}{|\partial_x h|} + \mathcal{O}(\varepsilon^2), \tag{9}$$

which yields the following boundary condition:

$$\partial_y u + \mathcal{S}^{xy} = 0 \quad \text{at} \quad y = h. \tag{10}$$

Integrating now Eq. (7a) with respect to $y$ and applying the boundary conditions in Eqs. (3) and (10), we obtain the following expression for the streamwise velocity:

$$u = \left(\frac{y^2}{2} - yh\right)\partial_x(p + \Pi) - \int_0^y \mathcal{S}^{xy}(y')\,\mathrm{d}y'. \tag{11}$$

Substituting Eq. (11) into the kinematic boundary condition Eq. (5), yields the time-evolution equation for the thin-film height:

$$\partial_t h = \partial_x\left[\frac{h^3}{3}\partial_x\left(\phi'(h) - \gamma\,\partial_x^2 h\right) + \int_0^h\int_0^y \mathcal{S}^{xy}(y')\,\mathrm{d}y'\,\mathrm{d}y\right]. \tag{12}$$

To simplify the fluctuating term, we can integrate the nested integral by parts and define $\mathcal{S}(\mathbf{r};t) = \mathcal{S}^{xy}(\mathbf{r};t)$, so that we finally obtain the central equation of this work:

$$\partial_t h = \partial_x \left[ \frac{h^3}{3} \partial_x \left( \phi'(h) - \gamma\, \partial_x^2 h \right) + \int_0^h (h-y)\mathcal{S}(y)\,\mathrm{d}y \right], \tag{13}$$

where the correlation structure of the noise is given by:

$$\mathbb{E}(\mathcal{S}(x,y;t)\mathcal{S}(x',y';t')) = 2\,T\,q_x(x-x')q_y(y-y')\delta(t-t'). \tag{14}$$

Obviously, Eq. (13) shows exactly the same structure as the SPDE derived by Grün et al. [37], as we have closely followed their derivation. But at the same time there is a remarkable difference in that we are not imposing so far any specific structure for the correlators $q_x$ and $q_y$, which in the study of Grün et al. [37] are imposed to be Dirac's delta. Indeed we will study later a different limit, where the noise is perfectly correlated along the cross-stream direction, which as already noted in the Introduction it is more physically meaningful.

As also already emphasised in the Introduction, the main advantage of this approach to the thin-film dynamics is that the fluctuations are derived ab initio. That sets aside any controversy regarding the best noise to represent real fluctuations, as the noise is naturally derived from first principles within the self-consistent framework of FH. However, the fluctuating term in Eq. (13) is too convoluted for practical purposes and, also, against the spirit of the long-wave approximation to eliminate the dependence in the cross-stream direction since it still shows explicitly a dependency in this direction. In Sect. 4 we will discuss how to derive an equivalent dynamics by proposing a much simpler version of the noise that is more convenient for analytical and numerical purposes. Despite the formal derivation presented here, and later in Sect. 4, the stochastic thin-film model studied in this work does not fit in any existent theory for SPDEs, e.g. Hairer's theory on regularity structures [38], at least for now. The mathematical challenge of giving a mathematically rigorous meaning to this kind of equations is, obviously, far beyond the original scope of this work. Nevertheless, any progress on such an area of research will have an important impact in many statistical-mechanical applications, where this type of SPDE appears.

## 3 Deterministic Dynamics

### 3.1 Gradient Flow Structure

In this section we will demonstrate the gradient-flow structure of the deterministic part of the SPDE (13). To do that, we need to show that there exists an effective energy functional $\mathcal{H}(h)$ such that the deterministic part of the equation can be rewritten as the gradient of that functional with a certain metric. If we consider the functional:

$$\mathcal{H}(h) = \int_0^L \phi(h) + \frac{\gamma}{2}|\partial_x h|^2\,\mathrm{d}x, \tag{15}$$

where $L$ is the length of the domain in the $x$-direction, Eq. (13) can be rewritten as:

$$\partial_t h = \partial_x \left[ \frac{h^3}{3} \partial_x \left( \frac{\delta \mathcal{H}}{\delta h} \right) + \int_0^h (h-y)\mathcal{S}\,\mathrm{d}y \right], \tag{16}$$

with $\frac{\delta}{\delta h}$ representing the functional derivative. Neglecting the noise for the time being we have that

$$\partial_t h = \partial_x \left[ \frac{h^3}{3} \partial_x \left( \frac{\delta \mathcal{H}}{\delta h} \right) \right] =: g_h^{-1} D\mathcal{H}(h) = \nabla \mathcal{H}(h), \tag{17}$$

which can be given a geometric interpretation as gradient flow on the energy surface in $h$-space with a metric of $g_h^{-1} = \partial_x M(h) \partial_x$, where $M = h^3/3$ is the effective mobility, and $D\mathcal{H} := \frac{\delta \mathcal{H}}{\delta h}$. Indeed, in the particular case of $\phi(h) = 0$ Eq. (17) belongs to a more general family of fourth-order degenerate PDEs of the form

$$\partial_t h = \partial_x \left( M(h) \partial_x^3 h \right), \tag{18}$$

which models the dynamics of different physical systems depending on the choice of the "mobility" $M(h)$. For $M(h) = h^3 + \alpha h$, the equation describes the spread of a small viscous drop [35], and in the case of $M(h) = |h|$ it describes the behaviour of a thin neck of fluid in a Hele-Shaw cell [9] (see also the discussion in Ref. [11]). Similar equations are also obtained as the mean-field limits of interacting particle systems, with the resulting PDEs governing the evolution of the density of the system [1,6,7,16,23,45,46]. It is worth mentioning that a gradient-flow structure is not a common property, however this feature is also shared with another well-known equation, the Derrida–Lebowitz–Speer–Spohn (DLSS) equation [20]. The DLSS equation is also a fourth-order gradient flow which was originally derived for the study of interface fluctuations in a two-dimensional spin system. As noticed by Jüngel and Matthes [42], the common structure shared by both Eq. (18) and the DLSS equation yields similar analytical difficulties in proving the positivity or non-negativity of solutions.

Focusing only on the deterministic part of Eq. (16), i.e. without the noise term and setting $\phi(h) = 0$, the resulting deterministic equation (17) is also conservative. This means that if the system has initially a given mass given by $c = \int_0^L h(x; t = 0)\, \mathrm{d}x$, then the mass will remain constant over time. Moreover, for non-negative initial conditions, the positivity of the solution is preserved [3].

Let us now consider the time derivative of the energy functional which will give us information about the behaviour of the energy function over time, for $\phi(h) = 0$. For this purpose, let us assume periodic boundary conditions in $x$, hence:

$$\frac{\mathrm{d}\mathcal{H}}{\mathrm{d}t} = \int_0^L \frac{\delta \mathcal{H}}{\delta h} \partial_t h \, \mathrm{d}x \tag{19}$$

$$= \int_0^L -\gamma\, \partial_x h\, \partial_x \left( \frac{h^3}{3} \partial_x \left( -\gamma \partial_x^2 h \right) \right) \mathrm{d}x \tag{20}$$

$$= -\int_0^L \gamma^2 \frac{h^3}{3} (\partial_x^3 h)^2 \, \mathrm{d}x < 0, \quad \forall h \neq \frac{c}{L}, \tag{21}$$

with $c$ being the mass of the initial condition, as mentioned above. The sign of the temporal derivative of the energy functional informs us about the dissipative nature of the functional for all acceptable film heights (i.e. with $h \geq 0$ and mass $c$) except for $h^* = c/L$. Also, the functional $\mathcal{H}$ is bounded from below by 0 and is continuous. Thus, $\mathcal{H}$ satisfies the conditions required for it to be a Lyapunov functional [70]. This implies that $h^* = \frac{c}{L}$ is the only steady state for $\phi(h) = 0$ and that it is globally attractive.

In the presence of a non-zero interaction potential $\phi$, we can explicitly compute the first and second variations of $\mathcal{H}$:

$$\delta\mathcal{H}(h,\eta) = \int_0^L (-\gamma\partial_x^2 h + \phi'(h))\eta\,\mathrm{d}x, \tag{22a}$$

$$\delta^2\mathcal{H}(h,\eta) = \int_0^L \gamma(\partial_x\eta)^2 + \eta^2\phi''(h)\,\mathrm{d}x, \tag{22b}$$

where $\eta$ belongs to a suitably defined space of admissible variations. It can be easily shown that any critical point of $\mathcal{H}$ (determined by $\delta\mathcal{H}(h,\eta) = 0$, $\forall\eta$) is a stationary solution of the deterministic PDE (17). What is more, given a critical point of $\mathcal{H}$ we can solve a variational problem in $\eta$ for the functional $\delta^2\mathcal{H}(h,\eta)$. Assuming $\eta$ is sufficiently smooth, we obtain the following identity:

$$\gamma\,\partial_x^2\eta = \eta\,\phi''(h). \tag{23}$$

Multiplying both sides of Eq. (23) by $\eta$, it is straightforward to get:

$$\eta^2\phi''(h) = \gamma\,\partial_x^2\left(\frac{\eta^2}{2}\right) - \gamma\,(\partial_x\eta)^2. \tag{24}$$

Substituting this result into Eq. (22b) and assuming periodic boundary conditions in $x$, we get that $\delta^2\mathcal{H} = 0$ for all critical points $\eta$ (and, thus, by extension for all minimisers) of this functional. This results in $\delta^2\mathcal{H}(h,\eta) \geq 0$ for all critical points $h$ of $\mathcal{H}$ and all admissible $\eta$. Hence, the critical points satisfy at least the necessary conditions for being minimisers of the functional $\mathcal{H}$.

### 3.2 Linear Stability Analysis of the Uniform State

In what follows we study the linear stability of an initial uniform state with constant height. To do that, we can linearise Eq. (17) about the state $h^* = 1$ with mass $c = L$, by inserting perturbations of the form:

$$h = h^* + \epsilon g, \tag{25a}$$

$$\int_0^L g(x)\,\mathrm{d}x = 0, \tag{25b}$$

with $g(x) > -1,\ \forall x \in [0, L]$.

Let us first consider a system with zero interaction potential, $\phi(h) = 0$. Retaining only the terms of $\mathcal{O}(\epsilon)$ in the equation, the resulting linearised operator $\mathscr{N}$ can be written as:

$$\mathscr{N} g = -\frac{\gamma}{3}\partial_x^4 g. \tag{26}$$

Considering again periodic boundary conditions, the eigenvalues $\lambda_n$ and eigenfunctions $g_n$ of the linearised operator are the following:

$$\lambda_n = -\gamma\frac{4\pi^4 n^4}{L^4}, \tag{27}$$

$$g_n = \sqrt{\frac{1}{L}}\exp\left(\mathrm{i}\frac{2\pi n x}{L}\right), \tag{28}$$

for all $n \in \mathbb{Z}\backslash\{0\}$. As can be immediately seen, the eigenvalues $\lambda_n$ are strictly negative. As a result, the system is unconditionally stable to sufficiently small perturbations.

Let us consider now the interesting case of a non-zero interaction potential that can be written as the sum of a non-negative convex and a concave function of the height. Consider, e.g., the case of $\phi(h) = \alpha h^{-3} - \beta h^{-2}$ with $\alpha, \beta \in \mathbb{R}^+$. The physical basis of this form goes back to the work by Derjaguin and Framkin, while using elements from the statistical mechanics of classical fluids, namely DFT, it can be shown that the (local) Derjaguin–Framkin disjoining pressure is an asymptote to that obtained from DFT as the distance of the chemical potential from its saturation value vanishes [79]. The linearised operator can then be written as follows,

$$\mathcal{N} g = (4\alpha - 2\beta)\, \partial_x^2 g - \frac{\gamma}{3}\, \partial_x^4 g. \tag{29}$$

For $x \in [0, \infty)$, we assume the factorisation $g(x; t) = e^{ikx} G(t)$, with $k = (2\pi/L) \in \mathbb{R}$ the wavenumber and $L$ the wavelength. This yields

$$G(t) = \exp\left\{\left[(2\beta - 4\alpha)k^2 - \frac{\gamma}{3}k^4\right]t\right\}. \tag{30}$$

Therefore, one branch of the neutral curve is given by $k = 0$ and the other is given by

$$(2\beta - 4\alpha) = \frac{\gamma}{3}k^2, \tag{31a}$$

$$|k_{\text{cr}}| = \sqrt{\frac{6(\beta - 2\alpha)}{\gamma}}, \tag{31b}$$

which defines the minimum wavelength (and, hence, domain length) for instability:

$$L_{\text{cr}} = \frac{2\pi}{k_{\text{cr}}} = 2\pi\sqrt{\frac{\gamma}{6(\beta - 2\alpha)}}. \tag{32}$$

Now, we can readily compute the wavenumber associated with the maximum growth rate, $k_{\max}$, and the corresponding domain length, $L_{\max}$,

$$k_{\max} = \frac{1}{\sqrt{2}}k_{\text{cr}}, \tag{33a}$$

$$L_{\max} = \sqrt{2}L_{\text{cr}}, \tag{33b}$$

From (31b) it is straightforward that the condition for the system to be unstable is $\beta > 2\alpha$. These results are summarised in Fig. 2.

## 4 Equivalent Stochastic Dynamics

Having characterised the linear stability conditions for the deterministic part of the stochastic thin-film equation derived in Sect. 2, we turn to the effect of fluctuations. As repeatedly mentioned so far, the structure of the noise in Eq. (13) is not yet convenient for practical purposes. That is why there is a need for a statistically equivalent equation involving a simpler noise term. Particularly, Grün et al. [37] have shown that there exists an equivalence in law between the above following two SPDEs:

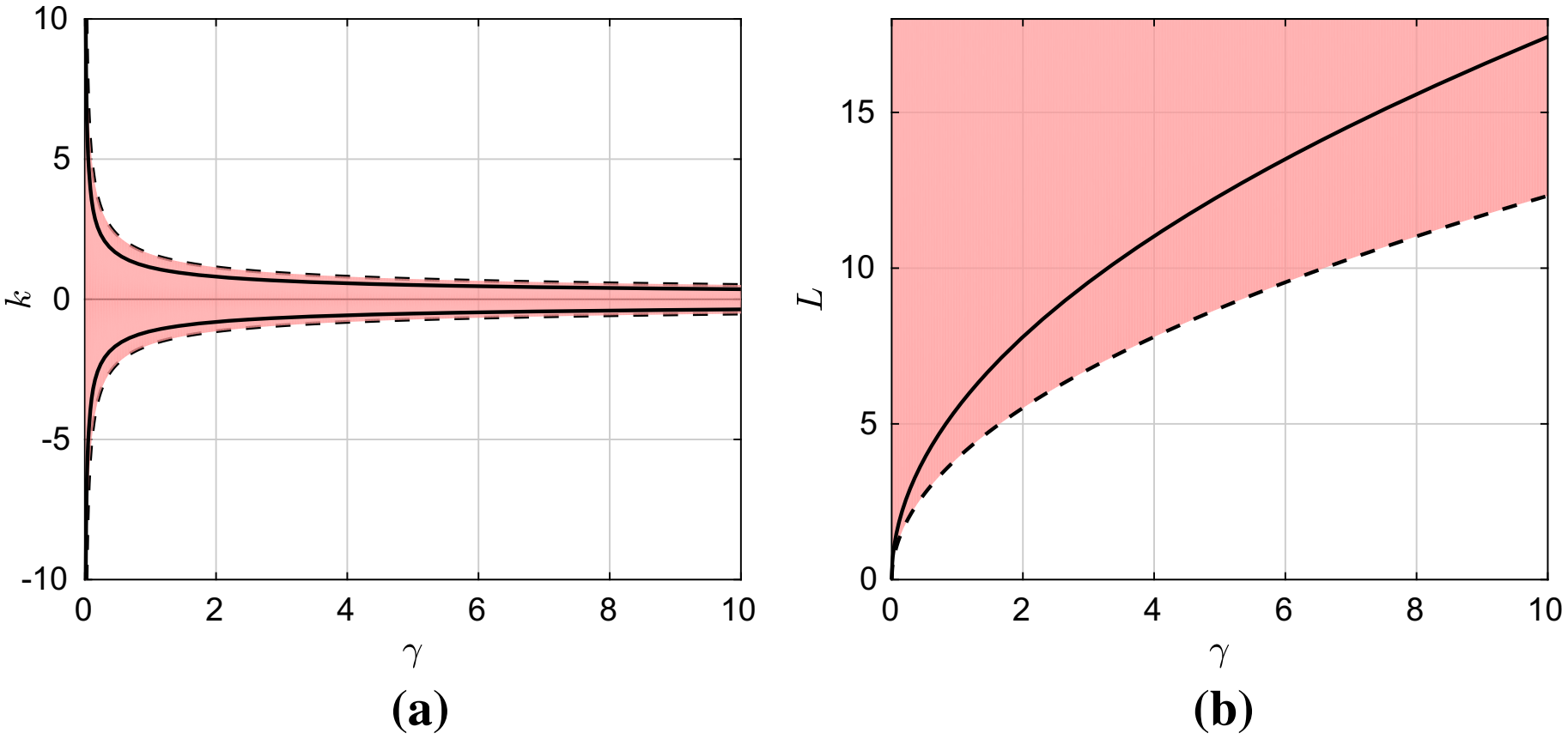

**Fig. 2** Linear stability of the uniform state for the interaction potential $\phi = \alpha h^{-3} - \beta h^{-2}$ with $\alpha = \frac{1}{30}$ and $\beta = \frac{1}{2}$ as a function of the surface tension coefficient, $\gamma$. **a** The stability map in terms of the $k$ and $\gamma$ values, whereas **b** the same on the $L$–$\gamma$ plane. Shaded red regions correspond to the parameter values for which the system is unstable. Dashed lines represent the critical stability curve given by Eqs. (31b) and (32). Solid lines represent the values associated with the maximum growth rate, Eq. (33)

$$\partial_t h = \partial_x \left[ \frac{h^3}{3} \partial_x \left( \frac{\delta \mathcal{H}}{\delta h} \right) + \int_0^h (h - y) \mathcal{S}(y) \, \mathrm{d}y \right], \tag{34a}$$

$$\partial_t h = \partial_x \left[ \frac{h^3}{3} \partial_x \left( \frac{\delta \mathcal{H}}{\delta h} \right) + \sqrt{\frac{h^3}{3}} \mathcal{N} \right], \tag{34b}$$

where $\mathcal{S}$ and $\mathcal{N}$ are zero-mean spatiotemporal noise functions, defined by their correlation structure:

$$\mathbb{E}(\mathcal{S}(x, y, t)\mathcal{S}(x', y', t')) = 2T\delta(t - t')q_x(x - x')q_y(y - y'), \tag{35a}$$

$$\mathbb{E}(\mathcal{N}(x, t)\mathcal{N}(x', t')) = 2T\delta(t - t')\delta(x - x'). \tag{35b}$$

Specifically, the equivalence is obtained by using a finite-difference discretisation of (34a) and, then, computing the related Kramers-Moyal coefficients to show that they both have associated the same Fokker–Planck equation (for delta-correlated noise in the $x$ and $y$ directions). In what follows, we use a finite-difference discretisation only to show that the simplified SPDE satisfies the detailed-balance condition (which is a requirement from thermodynamics [21]). This is done in Sect. 4.1. After that, in Sect. 4.2, we demonstrate that there exists an equivalency between the two SPDEs using a $Q$-Wiener representation of the noise function, $\mathcal{S}$. The reader should note that a similar cubic multiplicative noise is derived in Refs. [16] and [46] starting from a particle description of the system. We note that one would thus expect this equivalent SPDE to be obtained, at least in principle, as the mean field limit of a system of interacting stochastic processes.

### 4.1 Detailed-Balance Condition

In the original work of Grün et al. [37], it was stated without proof that the simplified SPDE in Eq. (34b) satisfies the detailed-balance condition. In this section we present a proof of this result. Our proof outlined below is based on a spatial discretisation of a more general version of Eq. (34b) with the multiplicative noise given by $\nu h^n$. In the end, we find a general condition that the noise must fulfil based on the Fokker–Planck equation associated with the resultant set of SDEs.

Let us consider the SPDE:

$$\partial_t h = \partial_x \left[ \frac{h^3}{3} \partial_x \left( \frac{\delta \mathcal{H}}{\delta h} \right) + \nu h^n \mathcal{N} \right], \tag{36}$$

where $\nu, n \in \mathbb{R}$ are free parameters for the moment, $\mathcal{H}$ being the energy functional given in Eq. (15). Let us also consider a finite difference discretisation of the $x$-axis such that $x_\ell = \ell\, a$, with $a \in \mathbb{R}_+$ the grid-spacing length and $\ell = 0, \dots, d-1$ the grid index. In that case, we have that:

$$\mathbf{h}(t) = [h_\ell]^\top_{h_\ell = h(\ell a; t)} = [h(0;t), h(a;t), \dots, h((d-2)a;t), h((d-1)a;t)]^\top \in \mathbb{R}^d. \tag{37}$$

Using now the differentiation matrix for a central-difference scheme $\mathbf{A} \in \mathbb{R}^{d\times d}$, we can rewrite Eq. (36) as follows:

$$\partial_t \mathbf{h}(t) = \mathbf{A}\left( \mathbf{j}(t) + \mathbf{K}(t)\boldsymbol{\xi}(t) \right), \tag{38}$$

where we have also introduced the discretised version of the drift, $\mathbf{j} = (j_\ell) \in \mathbb{R}^d$, and of the noise intensity, given by the diagonal matrix $\mathbf{K} = (K_{\ell\ell}) \in \mathbb{R}^{d\times d}$:

$$j_\ell(t) = \frac{h_\ell^3(t)}{3} \left[ \mathbf{A} \left( \frac{1}{a} \nabla_{\mathbf{h}} \tilde{\mathcal{H}}(t) \right) \right]_\ell, \tag{39a}$$

$$K_{\ell\ell}(t) = \nu h_\ell^n(t). \tag{39b}$$

We also define the diagonal matrix, $\mathbf{K}' = (K'_{\ell\ell}) \in \mathbb{R}^{d\times d}$ as follows,

$$K'_{\ell\ell}(t) = \frac{1}{\sqrt{3}} h_\ell^{3/2}(t). \tag{39c}$$

In the above expressions, the discretised version $\tilde{\mathcal{H}}$ of the functional $\mathcal{H}$ is defined by the Riemman sum $\tilde{\mathcal{H}} = a \sum_l \left( \phi'(h_l) - \frac{\gamma}{2} (\mathbf{A}\mathbf{h})_l^2 \right)$ and $\boldsymbol{\xi} = (\xi_1, \dots, \xi_d)^\top$ is a Gaussian white noise with the following correlation structure:

$$\mathbb{E}(\xi_\ell(t)\xi_m(t')) = \frac{2T}{a} \delta_{\ell m} \delta(t - t'). \tag{40}$$

The adjoint of the infinitesimal generator, $\mathscr{L}^\dagger$, of the Markov semigroup associated with this SDE is then given by:

$$\mathscr{L}^\dagger \phi = \nabla_{\mathbf{h}} \cdot \left( (-\mathbf{A} \cdot \mathbf{j})\phi + \frac{T}{a} \nabla_{\mathbf{h}} \cdot (\Sigma \phi) \right), \tag{41}$$

with $\Sigma = \mathbf{AK}(\mathbf{AK})^T$. Thus, the Fokker–Planck equation governing the temporal evolution of the probability density function (PDF) to observe a certain state at a given time, $\rho(\mathbf{h}; t)$, can be written as:

$$\partial_t \rho = \mathscr{L}^\dagger \rho = \nabla_{\mathbf{h}} \cdot \left( -\mathbf{A}\mathbf{j}\rho - \frac{T}{a} (\mathbf{AKK}^T \mathbf{A} \nabla_{\mathbf{h}} \rho) \right), \tag{42}$$

with the initial condition $\rho_0 = \rho(0)$. Here we have used the fact that $\mathbf{A}$ is antisymmetric and that $\nabla_{\mathbf{h}}(\mathbf{AKK}^T\mathbf{A}) = 0$. Inserting the definition of the discretised drift given in Eq. (39a) into

Eq. (42), we obtain,

$$\partial_t æ = \nabla_{\mathbf{h}} \cdot \left( \left( -\frac{1}{a} \mathbf{A}\mathbf{K}'\mathbf{K}'^T \mathbf{A} \nabla_{\mathbf{h}} \tilde{\mathcal{H}} - \frac{T}{a} \mathbf{A}\mathbf{K}\mathbf{K}^T \mathbf{A} \nabla_{\mathbf{h}} \right) \rho \right) \tag{43}$$
$$= -\frac{1}{a} \nabla_{\mathbf{h}} \cdot J(\rho),$$

with:

$$J(\rho) = \left( \mathbf{A}\mathbf{K}'\mathbf{K}'^T \mathbf{A} \nabla_{\mathbf{h}} \tilde{\mathcal{H}} + T \mathbf{A}\mathbf{K}\mathbf{K}^T \mathbf{A} \nabla_{\mathbf{h}} \right) \rho. \tag{44}$$

If we now set $\rho = \rho_s$ to be the Gibbs measure associated with $\tilde{\mathcal{H}(\mathbf{h})}$, i.e., $\rho_s = \frac{1}{Z} e^{-\frac{1}{T} \tilde{\mathcal{H}(\mathbf{h})}}$ for some normalisation constant $Z$, we obtain:

$$J(\rho_s) = \left( \mathbf{A} \left( \mathbf{K}'\mathbf{K}'^T - \mathbf{K}\mathbf{K}^T \right) \mathbf{A} \nabla_{\mathbf{h}} \tilde{\mathcal{H}} \right) \rho_s = \Gamma(\mathbf{h}) \rho_s, \tag{45}$$

Now, we need to show that if $\mathbf{K}'\mathbf{K}'^T - \mathbf{K}\mathbf{K}^T \neq 0, \forall \mathbf{h} \in \mathbb{R}^d$ there exists and admissible, $\mathbf{h} \in \mathbb{R}^d$ for which $\Gamma(\mathbf{h}) \neq 0$. It is sufficient to show this for only one component of the vector, $[\Gamma \mathbf{h}]_l$. To do this, we define the function, $f(h_\ell)$, as follows,

$$f(h_\ell) = \frac{h_\ell^3}{3} - \nu^2 h_\ell^{2n}. \tag{46}$$

Then, after some straightforward but lengthy calculations, we obtain an explicit expression for $[\Gamma(\mathbf{h})]_0$,

$$[\Gamma(\mathbf{h})]_0 = [-f(h_1) - f(h_{d-1})] \left[ 12\alpha h_0^{-5} - 6\beta h_0^{-4} - 2(h_2 - h_0) + 2(h_0 - h_{d-2}) \right]$$
$$+ f(h_1) \left[ 12\alpha h_2^{-5} - 6\beta h_2^{-4} - 2(h_4 - h_2) + 2(h_2 - h_0) \right]$$
$$+ f(h_{d-1}) \left[ 12\alpha h_{d-2}^{-5} - 6\beta h_{d-2}^{-4} - 2(h_0 - h_{d-2}) + 2(h_{d-2} - h_{d-4}) \right]. \tag{47}$$

We can now pick $h_0 = h_2 = h_{d-2} = h_{d-4} = \kappa/(2da)$, $h_4 = 2h_2$ for some $0 < \kappa < 1$ and pick $h_1 < 1/(ad)$ such that $f(h_1) \neq 0$. The other components of $\mathbf{h}$ can be chosen such that $\sum_\ell h_\ell a = 1$. This gives us $[\Gamma(h_0)] \neq 0$. Thus the flux $J(\rho_s)$ can be zero, $\forall h \in \mathbb{R}^d$, if and only if $\nu = \pm\frac{1}{\sqrt{3}}$ and $n = \frac{3}{2}$. That is, the Fokker–Planck equation satisfies the detailed-balance condition and, thus, the generator $\mathscr{L}$ is symmetric [59], if and only if the general noise in Eq. (36) has exactly the same noise coefficient, $\nu$ and dependency on $h$, as proposed in Refs. [12,37].

This result is in agreement with the general fluctuation-dissipation relation for general stochastic gradient flows $dz = g^{-1}(z) D\mathcal{H}(z) dt + \sigma(z) d\mathcal{W}(t)$, with $K$ the metric, $\sigma$ an operator acting on $d\mathcal{W}(t)$ and $\mathcal{W}(t)$ a Brownian sheet, recently discussed by Reina and Zimmer [67]. Following these authors, for the MEP and the LD principles to be fulfilled concomitantly, the relationship $g^{-1} \propto \sigma\sigma^*$ must be satisfied. In our particular case, that means:

$$\partial_x \left( \frac{h^3}{3} \partial_x f(x) \right) \propto \partial_x \left( \nu^2 h^{2n} \partial_x f(x) \right), \tag{48}$$

whence, $\nu = \pm\frac{1}{\sqrt{3}}$ and $n = \frac{3}{2}$ if the proportionality constant is assumed to be unity. As can be seen, the detailed-balance condition is more restrictive as it imposes the proportionality constant without doubt.

In that case, the stationary process associated with the SDE in Eq. (38) is reversible, i.e. for every $T \in [0, \infty)$, $\mathbf{h}(t)$ and the time-reversed process $\mathbf{h}(T - t)$ have the same finite-dimensional distribution [59]. This means that, given any finite sequence of times $t_0 < t_1 < t_2 < \cdots < t_k = T$, and corresponding measurable subsets, $A_0, A_1, A_2, \ldots, A_k$, the following identity is true:

$$\mathbb{P}(\mathbf{h}(t_0) \in A_0, \ldots, \mathbf{h}(t_k) \in A_k) = \mathbb{P}(\mathbf{h}(T - t_0) \in A_0, \ldots, \mathbf{h}(T - t_k) \in A_k), \tag{49}$$

which means that, statistically, the stationary process is insensible to the time-arrow direction.

### 4.2 Representation of the Noise in Two Dimensions

We turn now our attention back to Eq. (34a) with the aim of representing the spatiotemporal fluctuations, $\mathcal{S}$, in a more convenient way, as we already mentioned before. When the noise term is finally represented as an infinite expansion in terms of independent real-valued Brownian motions, we will be able to impose the long correlation-length limit along the cross-stream direction, i.e. $l_y \to \infty$, and show that the noise in Eq. (34a) (hence, the SPDE itself) converges to the much simpler one-dimensional noise structure of Eq. (34b).

Consider a finite domain of length $L$ along the streamwise direction $x$, and height $h(x; t)$ along the wall-normal direction $y$, with periodic boundary conditions along $x$. Let $H$ be the Hilbert space of all square integrable functions on $[0, L] \times [0, h]$. Assume that for every time $t \in [0, \infty)$, $\mathcal{S}$ takes values in $H$. It is known that a spatiotemporal noise process can be written as the formal time derivative of a $Q$-Wiener process, $\mathcal{W}$, such that [15,75]:

$$\mathcal{S} = \frac{\mathrm{d}}{\mathrm{d}t}\mathcal{W}, \tag{50}$$

where $\mathcal{W}$ is an infinite-dimensional zero-mean Gaussian process which takes values in $H$ and is defined entirely by the covariance operator $Q$, which is symmetric, positive and of finite trace. Consider the following result [64]:

**Proposition 1** *Let $Q \in L(H)$[1] be a symmetric, non-negative operator with* $\mathrm{tr}(Q) < \infty$. *Further, we assume that $g_k \in H$, $k \in \mathbb{Z}$ is an orthonormal basis consisting of the eigenvectors of $Q$ with eigenvalues $\lambda_k$. Then an $H$-valued random variable, $\mathcal{W}$, is Gaussian or $Q$-Wiener if and only if,*

$$\mathcal{W} = \sum_{k=-\infty}^{\infty} \sqrt{\lambda_k}\, g_k \beta_k(t), \tag{51}$$

*where $\beta_k(t)$ are independent real-valued Brownian motions and where the convergence is in $L^2(\Omega, \mathcal{F}, \mathbb{P}; H)$, the space of Bochner-square integrable functions, $f : \Omega \to H$.*

Applying Prop. 1, we can obtain a formal representation for $\mathcal{S}$ as follows:

$$\mathcal{S}(x, y; t) = \sum_{k=-\infty}^{\infty} \sqrt{\lambda_k}\, g_k \dot{\beta}_k(t), \tag{52}$$

[1] $L(H)$ is the space of bounded linear operators on $H$.

where $\dot{\beta}_k : \Omega \times [0, \infty) \to \mathbb{R}$ are independent white-noise processes, i.e. zero-mean Gaussian processes with correlation determined by $\mathbb{E}(\dot{\beta}_k(t)\dot{\beta}_l(s)) = \delta_{kl}\delta(t-s)$, and $\lambda_k$ and $g_k$ are the eigenvalues and eigenvectors, respectively, of the operator $Q$, defined by its action on a field:

$$Qf(x, y) \equiv \int_0^L \int_0^h 2T q_x(x - x')q_y(y - y')f(x', y')\,\mathrm{d}x'\,\mathrm{d}y' \tag{53}$$

We propose the functions $q_x$ and $q_y$ to be:

$$q_x(x) = Z_x^{-1} \exp\left(-\frac{1}{2}\sin^2\left(\frac{\pi x}{L}\right)\left(\frac{L^2}{l_x^2}\right)\right), \tag{54}$$

$$q_y(y) = Z_y^{-1} \exp\left(-\frac{1}{2}\sin^2\left(\frac{\pi y}{h}\right)\left(\frac{h^2}{l_y^2}\right)\right), \tag{55}$$

where $l_x, l_y$ are the correlation lengths, and $Z_x,\ Z_y$ the normalisation constants, along $x$ and $y$, respectively. We set $Z_x$ such that $\int_{-L/2}^{L/2} q_x \mathrm{d}x = \sqrt{2T}$, while $Z_y$ is left undefined for the time being. $Q$ is symmetric and nonnegative, as expected. The trace is given by $\int_0^L \int_0^h 2T Z_x^{-1} Z_y^{-1}\,\mathrm{d}x\,\mathrm{d}y$ and is, thus, finite. To close the alternative representation of the noise we need to solve the eigenvalue problem:

$$Qg_k = \lambda_k g_k. \tag{56}$$

The structure of the operator $Q$ motivates us to write the eigenfunctions $g_k$ as the product of two functions $g_k(x, y) = X(x)Y(y)$, so that the eigenvalue problem is rewritten as:

$$\left(\int_0^L 2T q_x(x - x')X_m(x')\,\mathrm{d}x'\right)\left(\int_0^h q_y(y - y')Y_n(y')\,\mathrm{d}y'\right) = a_m b_n X_m Y_n. \tag{57}$$

The solution to this eigenvalue problem has the following eigenfunctions:

$$X_m = \begin{cases} \sqrt{\frac{2}{L}}\cos\left(\frac{2\pi m x}{L}\right), & m > 0, \\ \sqrt{\frac{2}{L}}\sin\left(\frac{2\pi m x}{L}\right), & m < 0, \\ \sqrt{\frac{1}{L}}, & m = 0, \end{cases} \quad Y_n = \begin{cases} \sqrt{\frac{2}{h}}\cos\left(\frac{2\pi n y}{h}\right), & n > 0, \\ \sqrt{\frac{2}{h}}\sin\left(\frac{2\pi n y}{h}\right), & n < 0, \\ \sqrt{\frac{1}{h}}, & n = 0, \end{cases} \tag{58}$$

with the eigenvalues given by,

$$a_m = \begin{cases} 2T Z_x^{-1} \exp\left(-\frac{L^2}{4l_x^2}\right) \int_0^L \exp\left(\frac{L^2}{4l_x^2}\cos\left(\frac{2\pi z}{L}\right)\right)\cos\left(\frac{2\pi m z}{L}\right)\mathrm{d}z, & m \neq 0, \\ (2T)^{3/2} & m = 0, \end{cases} \tag{59a}$$

$$b_n = \begin{cases} Z_y^{-1} \exp\left(-\frac{h^2}{4l_y^2}\right) \int_0^h \exp\left(\frac{h^2}{4l_y^2}\cos\left(\frac{2\pi z}{h}\right)\right)\cos\left(\frac{2\pi n z}{h}\right)\mathrm{d}z, & n \neq 0, \\ \int_0^h q_y\,\mathrm{d}y & n = 0. \end{cases} \tag{59b}$$

At this stage, we require that $b_0$ must be constant and independent of $l_y$. Thus, the normalisation constant $Z_y$ must be defined as,

$$Z_y = Z_y(l_y) = \frac{1}{b_0}\int_0^h \exp\left(-\frac{1}{2}\sin^2\left(\frac{\pi y}{h}\right)\left(\frac{h}{l_y}\right)^2\right)\mathrm{d}y, \quad b_0 \in \mathbb{R}^+. \tag{60}$$

Finally, we have all the ingredients for the easier representation of $\mathcal{W}$, and hence for $\mathcal{S}$ according to Eq. (52),

$$\mathcal{W} = \sum_{m=-\infty}^{\infty} \sum_{n=-\infty}^{\infty} \sqrt{a_m b_n} X_m Y_n \beta_{mn}(t), \tag{61}$$

together with the definitions given before in Eqs. (58)–(60) and with $\beta_{mn}(t)$ as independent 1D Brownian motions.

### 4.3 Long Correlation-Length Limit: Perfectly Correlated Noise Along the Wall-Normal Direction

As we mentioned at the beginning of Sect. 4.2, what is left to obtain Eq. (34b) is to take the limit of $\mathcal{W}$ when the correlation length goes to infinity. It is true that in the original Landau and Lifshitz derivation the correlations are considered to be Diracs in every direction, as is the case in many other fluctuating hydrodynamics derivations where the system under consideration is open (without boundaries) (e.g. [24]). Our approach on the other hand, is more "natural" motivated from physical considerations as we explained in the Introduction. But at the same time, there is no rigorous justification of what the autocorrelation function looks like for thin films in the presence of a wall and in that respect, our approach is effectively a hypothesis when it comes to modelling fluctuations. For this reason, we are considering the generalisation presented in this work. Thus,

$$\lim_{l_y \to \infty} \mathcal{W} = \lim_{l_y \to \infty} \sum_{m=-\infty}^{\infty} \sum_{n=-\infty}^{\infty} \sqrt{a_m b_n} X_m Y_n \beta_{mn}(t). \tag{62}$$

Consider now the behaviour of the eigenvalues, $b_n, \ \forall n \neq 0$, in the same limit:

$$\begin{aligned}
\lim_{l_y \to \infty} b_n &= \lim_{l_y \to \infty} Z_y^{-1} \exp\left(-\frac{h^2}{4l_y^2}\right) \int_0^h \exp\left(\frac{h^2}{4l_y^2} \cos\left(\frac{2\pi z}{h}\right)\right) \cos\left(\frac{2\pi n z}{h}\right) \mathrm{d}z \\
&= \left(\lim_{l_y \to \infty} Z_y^{-1}\right) \left(\lim_{l_y \to \infty} \exp\left(-\frac{h^2}{4l_y^2}\right)\right) \\
&\quad \times \left(\lim_{l_y \to \infty} \int_0^h \exp\left(\frac{h^2}{4l_y^2} \cos\left(\frac{2\pi z}{h}\right)\right) \cos\left(\frac{2\pi n z}{h}\right) \mathrm{d}z\right) \\
&= b_0 \frac{\lim_{l_y \to \infty} \int_0^h \exp\left(\frac{h^2}{4l_y^2} \cos\left(\frac{2\pi z}{h}\right)\right) \cos\left(\frac{2\pi n z}{h}\right) \mathrm{d}z}{\lim_{l_y \to \infty} \int_0^h \exp\left(\frac{h^2}{4l_y^2} \cos\left(\frac{2\pi z}{h}\right)\right) \mathrm{d}z}, \quad n \neq 0.
\end{aligned} \tag{63}$$

By dominated convergence we have,

$$\lim_{l_y \to \infty} b_n = 0, \quad n \neq 0. \tag{64}$$

Now consider the $Q$-Wiener process $\mathcal{W}'$ having the following representation,

$$\mathcal{W}' = \sum_{m=-\infty}^{\infty} \sqrt{b_0} Y_0 \sqrt{a_m} X_m \beta_{0m}(t). \tag{65}$$

Fix $t > 0$, and consider the sequence of random variables, $\mathcal{W}^{l_y}(t) : \Omega \to H$. Then, we have:

$$||\mathcal{W}'(t) - \mathcal{W}^{l_y}(t)||_{L^2} \leq \epsilon + \left( \int_\Omega \left\| \sum_{n=-N, n\neq 0}^{N} \sum_{m=-M}^{M} \sqrt{b_n a_m} Y_n X_m \beta_{mn}(w, t) \right\|_2^2 \mathrm{d}\omega \right)^{1/2}$$
$$\leq \epsilon + \sum_{n=-N, n\neq 0}^{N} \sum_{m=-M}^{M} \sqrt{b_n a_m} t, \tag{66}$$

for some $\epsilon$ which can be made arbitrarily small. Using this result and Eq. (64), we have $\mathcal{W} \to \mathcal{W}'$ as $l_y \to \infty$, in $L^2(\Omega; H)$. Thus, we can formally write that $\mathcal{S}^{l_y} \to \mathcal{D} = \frac{\mathrm{d}\mathcal{W}'}{\mathrm{d}t}$ as $l_y \to \infty$. Finally, inserting this representation into (34a) one gets:

$$\partial_t h = \partial_x \left( \frac{h^3}{3} \partial_x \left( \frac{\delta \mathcal{H}}{\delta h} \right) + \sum_{m=-\infty}^{\infty} \sqrt{a_m} X_m \int_0^h (h - y) \sqrt{b_0} Y_0 \mathrm{d}y \, \dot{\beta}_m \right). \tag{67}$$

Integrating the last term of the latter equation we eventually obtain the expression we were after:

$$\partial_t h = \partial_x \left( \frac{h^3}{3} \partial_x \left( \frac{\delta \mathcal{H}}{\delta h} \right) + \frac{\sqrt{b_0}}{2} \sqrt{h^3} \sum_{m=-\infty}^{\infty} \sqrt{a_m} X_m \dot{\beta}_m \right). \tag{68}$$

As can be seen, the infinite sum in this equation represents the spatiotemporal noise $\mathcal{N}$ of Eq. (34b). Thus, we can conclude that in the long correlation-length limit the SPDE (34a) converges to a $b_0$-parametrised family of SPDEs with a simpler noise structure. We can now select a particular value of $b_0$, which was left intentionally undefined before. The choice of that constant must be such that the resulting Fokker–Planck equation satisfies the detailed-balance condition with the invariant measure, $\rho_s$, as specified in the previous section. Under such circumstances, $b_0$ should be equal to $4\nu^2 = \frac{4}{3}$ and, therefore, the term multiplying the noise $\mathcal{N}$ will become $\pm \frac{h^{\frac{3}{2}}}{\sqrt{3}}$, as postulated by Davidovitch et al. [12] and Grün et al. [37]. It is also interesting to note that numerical computation of the eigenvalues $b_n$ reveals that they are distributed according to a discrete Gaussian distribution of the form (see Fig. 3):

$$b_n = b_0 \exp \left( - \left( \frac{\sqrt{2} n l_y}{h} \right)^2 \right), \quad n \in \mathbb{Z}. \tag{69}$$

Hence, choosing $b_0$ is equivalent to selecting the normalisation constant of the distribution. Computations of $b_n$ and their apparent distribution are shown in Fig. 3 for different values of the correlation-length $l_y$.

## 5 Numerical Experiments

Having shown the equivalency between the two equations (34a) and (34b), we perform detailed numerical experiments using a spectral method as an alternative to the finite-element discretisation used in the previous study by Grün et al. [37]. Spectral methods give us the opportunity of selecting, in a straightforward way, the frequency modes of the noise (following the spirit of the long-wave approximation), besides the convenience of smaller but denser matrices. SPDEs need to be discretised carefully, since different discretisations can converge

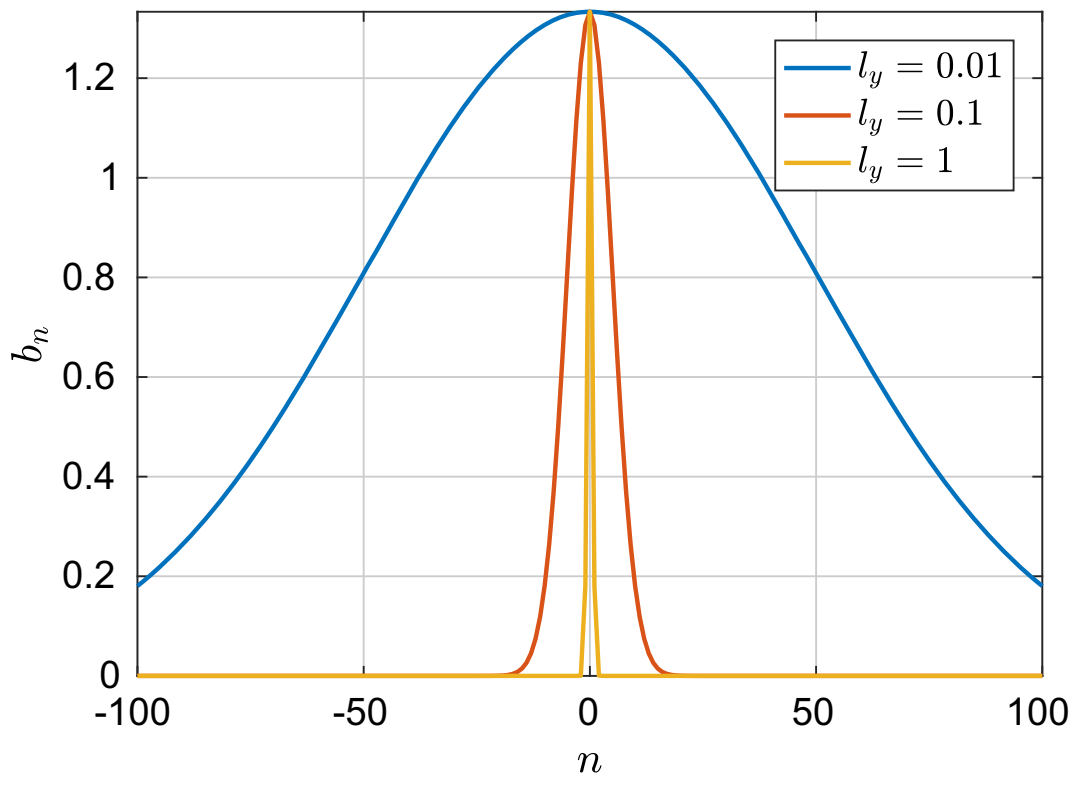

**Fig. 3** Distribution of the eigenvalues $b_n$ of the covariance operator $Q$ for different values of the correlation length along the wall-normal direction

to different limiting processes as the mesh size goes to zero. A good discussion about this issue is offered in Ref. [40], where discrepancies are shown for different finite-difference approximations of the stochastic Burgers equation.

In what follows, we provide a brief description of how our numerical method works and, then, proceed to validate it by computing statistics for the rupture times. Considering the SPDE in (34b), we know that there exists a trace-class, nonnegative covariance operator $Q'$ such that its eigenfunctions can be used to provide a spectral decomposition of the noise term $\mathcal{N}$. We argue that this set of eigenfunctions should form the natural setting to discretise the SPDE. Computing the solutions with such a discretisation, and observing a qualitative agreement with previous works [37], we intend to show that our discretisation method is correct. Nevertheless, a more exhaustive analysis of the sensitivity of the solutions on the number of points is yet to be done. This is however a question worth investigating in more detail in future studies, along with a rigorous examination of the numerical method, since the transition from a discrete solution to the actual continuous function that solves the SPDE is still a mathematical challenge.

## 5.1 Description of the Numerical Method

Consider the uniformly spaced partition $\{x_n\}_{-N}^{N} \subset [0, L]$ with spacing $\Delta x$. Again, we define the film height vector $\mathbf{h} \in \mathbb{R}^{2N+1}$, $\mathbf{h}_i = h(x_{i-N-1})$. Let $X_m$ and $a_m$ be the eigenfunctions and eigenvalues of $Q'$ respectively. Thus we can now define the spectral matrix, $C \in \mathbb{R}^{(2N+1)\times(2N+1)}$ such that $C_{ij} = X_{j-N-1}(x_{i-N-1})$ and the vector of eigenvalues, $\mathbf{a} \in \mathbb{R}^{2N+1}$ such that $\mathbf{a}_i = a_{i-N-1}$. This way, $Cx$ for any $x \in \mathbb{R}^{2N+1}$ is a Galerkin projection onto the space of eigenfunctions of $Q'$. If we choose $Q'$ as follows:

$$Q' f = \int_0^L 2T q_x(x - x') f(x')\,\mathrm{d}x', \quad \forall f \in L^2([0, L]), \tag{70}$$

then $X_m$ and $a_m$ are exactly those in (58) and (59b). This considerably simplifies the differentiation operation. We can use this to define the spectral differentiation matrices $D_n \in \mathbb{R}^{(2N+1)\times(2N+1)}$ that will be required for the numerical experiment. Let $B, A \in \mathbb{R}^{(2N+1)\times(2N+1)}$ be anti-diagonal and diagonal matrices, respectively, such that $B_{i,2N+2-i} = \mathrm{sign}(N+1-i)$ and $A_{ii} = |\frac{2\pi i-N-1}{L}|$. Then, we set $D_n = C^{-1}(AB)^n C$ so that $D_n\mathbf{h}$ amounts to an approximation of the $n^{\text{th}}$-derivative of $h$. We can now write down a nonlinear SDE with multiplicative noise which approximates the solution of the SPDE:

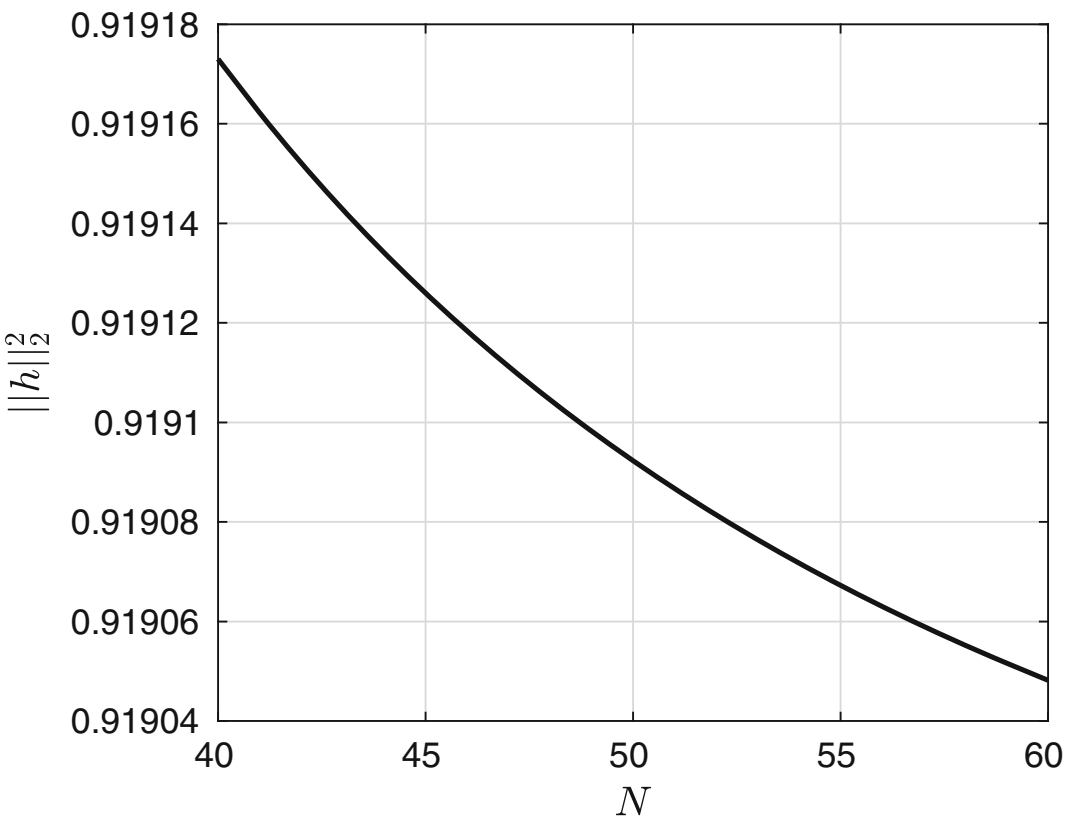

**Fig. 4** The $L^2$-norm of the film height averaged over 60 sample paths plotted against number of approximating modes, $N$ for a fixed time $\sqrt{2T} = 0.01$

$$\mathrm{d}\mathbf{h} = b(\mathbf{h})\mathrm{d}t + \sigma(\mathbf{h})\mathrm{d}W_t, \tag{71}$$

where:

$$b = D_1\left(\frac{\mathbf{h}^3}{3} D_1\left(\phi'(\mathbf{h}) - \gamma D_2\mathbf{h}\right)\right) \in \mathbb{R}^{2N+1}, \tag{72a}$$

$$\sigma = D_1\left(\left(I\mathbf{y}^T\sqrt{\frac{\mathbf{h}^3}{3}}\right) C(I\mathbf{y}^T\mathbf{a})\right) \in \mathbb{R}^{(2N+1)\times(2N+1)}, \tag{72b}$$

are the drift vector and the square root of the diffusion matrix, respectively, with $\mathbf{y} \in \mathbb{R}^{2N+1}$ defined as $\mathbf{y}_i = 1$.

Having the approximating SDE, all that is left is to choose an appropriate time-integration scheme. Since the deterministic problem is stiff, it is more appropriate and reliable to use two separate time-steps $\Delta t_s$ and $\Delta t_d < \Delta t_s$, for the stochastic and deterministic parts of the SDE, respectively. Consider the uniform partition $t_n \subset [0, T]$ with spacing $\Delta t_s$. Then, for each interval, $[t_i, t_i + \Delta t_s]$ we integrate over time the deterministic equation, $\mathrm{d}\mathbf{h} = b(\mathbf{h})\mathrm{d}t$ with $\mathbf{h}_{t_i}$ as the initial condition using MATLAB's `ode15s` stiff explicit multistep solver to obtain ${\mathbf{h}^*}_{t_i+\Delta t_s}$, where the solver selects $\Delta t_d$ adaptively. We then add the stochastic component using the Milstein scheme, i.e. $\mathbf{h}_{t_i+\Delta t_s} = {\mathbf{h}^*}_{t_i+\Delta t_s} + \sqrt{\Delta t_s}\sigma\xi + \frac{\Delta t_s}{2}\sigma\sigma^T(\xi^2 - 1)$, where $\xi$ is a vector of independent and identically distributed (iid) Gaussian random variables. As regards to the question of whether this numerical scheme satisfies a discrete version of the fluctuation–dissipation theorem we remark that it is not straightforward to define properly non-Gaussian Gibbs measures in infinite dimensions for SPDEs and thus it is hard to make the notion of such a theorem rigorous for the approximating SDE. We direct the reader to the following references for more details on fluctuation-dissipation theorems for SPDEs [4,5,17,39]. To conclude our discussion of the numerical scheme we include Fig. 4, showing convergence of the scheme in grid-size/number of modes.

## 5.2 Simulations

Here we perform numerical simulations of the deterministic and the stochastic thin-film equations. We know from our analysis in Sect. 3.2 that the uniform solution is linearly unstable. This raises the question of whether or not there exist other stationary solutions to the thin-film equation. Choosing the domain length to be an integer multiple of $L_{\max}$ and time-integrating with the initial condition $h_0 = 1 + \epsilon \sin(k_{\max} x)$ for $\epsilon \ll 1$, the system converges to a second stationary solution, as is evident in Fig. 5. The second stationary solution has a cluster-like structure and corresponds to the film rupturing and breaking up into droplets. This behaviour resembles that seen in systems of interacting particles in statistical mechanics where the uniform distribution destabilises and gives rise to molecular clusters (see, e.g., Refs. [14] and [52]) but also in systems with state transitions induced by thermal fluctuations (e.g. Ref. [62]).

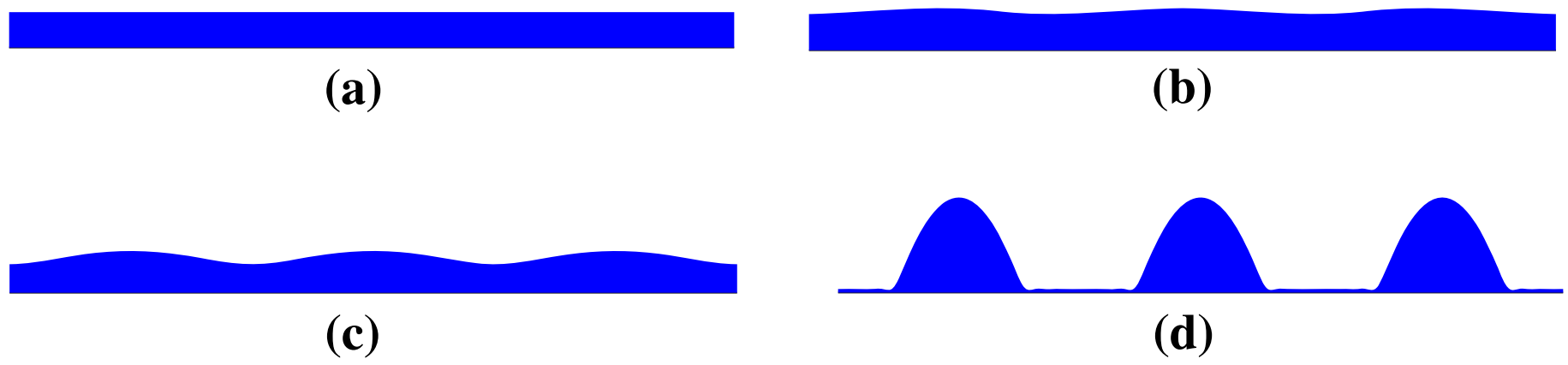

**Fig. 5** Deterministic evolution of a thin film under the influence of the disjoining potential, $\phi(h) = \frac{h^{-3}}{30} - \frac{h^{-2}}{2}$. The sequence must be understood in the following order (**a**)→(**b**)→(**c**)→(**d**). The uniform solution destabilises and then converges to a second stationary solution which has a cluster-like structure, i.e. the film ruptures and breaks up into droplets

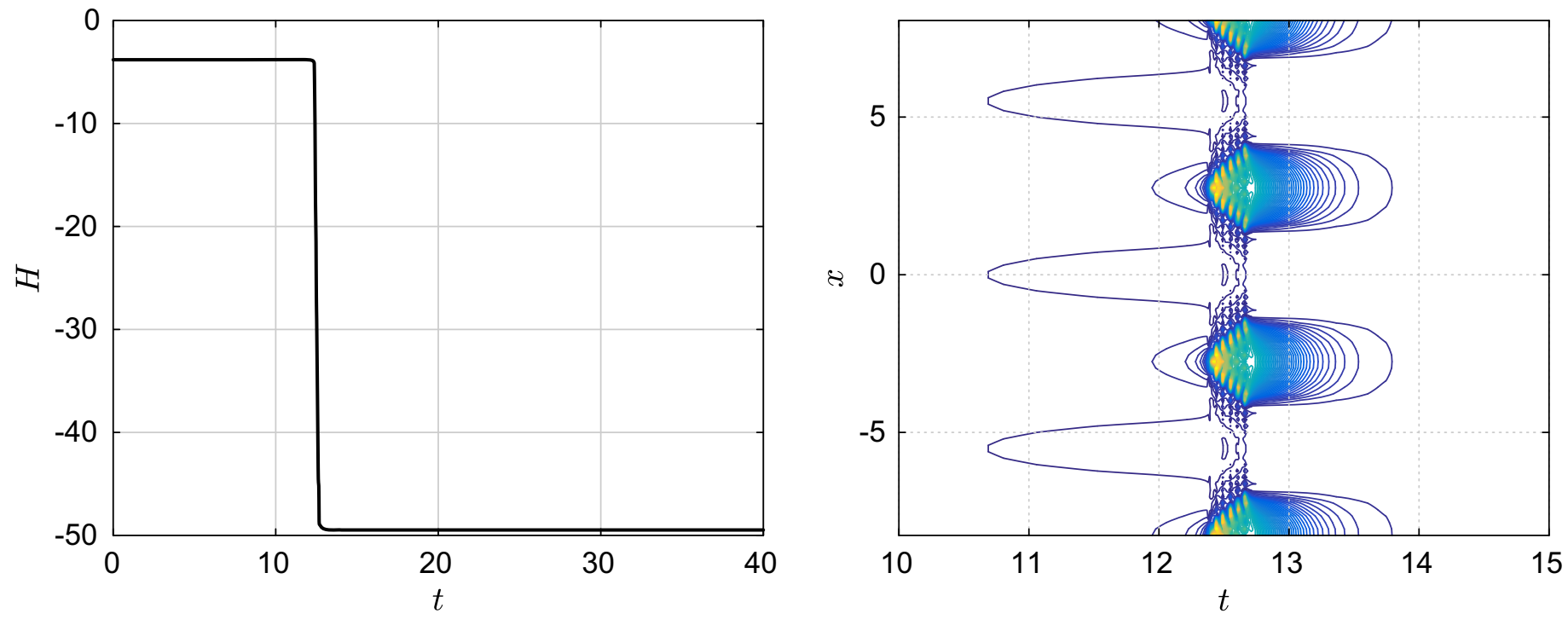

**Fig. 6** Evolution of the energy functional $\mathcal{H}$ as a function of time (left), showing two minima: the uniform and droplet states, and of its functional derivative $\frac{\delta \mathcal{H}}{\delta h}$ near the rupture time (right), where the two white regions represent values of $x$ and $t$ for which the functional derivative is zero

To ensure that there is no transient stationary solution between the uniform state and the droplet state, we have also computed the values of the free energy functional $\mathcal{H}$ as a function of time (see Fig. 6 (left)) and the functional derivative $\frac{\delta \mathcal{H}}{\delta h}$ near the rupture time (see Fig 6 (right)). As can be seen, the functional derivative is strictly non-zero between the two states, which indeed implies there is no transient intermediate solution. This calculation is completely necessary to know whether the system has to overcome an energy "barrier" during the breakup process, which then would comprise a nucleation event.

The inclusion of noise does not seem to induce a qualitative difference onto the behaviour of the thin film (see Fig. 7). The transitions from fluctuations about the uniform state to fluctuations about the droplet state can be seen in Fig. 7. To study the effect of noise on the thin-film dynamics we vary the value of the noise intensity and obtain statistics for the rupture time $t_r$, which we define as:

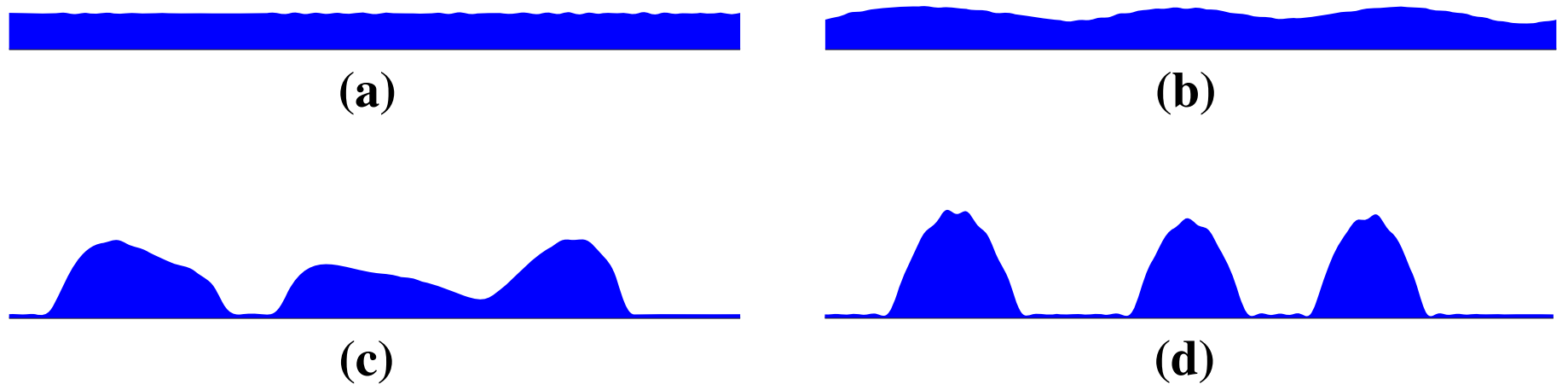

**Fig. 7** Temporal evolution of the thin film under the influence of the disjoining potential, $\phi(h) = \frac{h^{-3}}{30} - \frac{h^{-2}}{2}$ for $\sqrt{2T} = 0.01$ and $l_x = 0$. The sequence is organised in time as follows (**a**)→(**b**)→(**c**)→(**d**)

**Fig. 8** Single-path variation of the minimum film height near the rupture time with $\sqrt{2T} = 0.01$ and $l_x = 0$

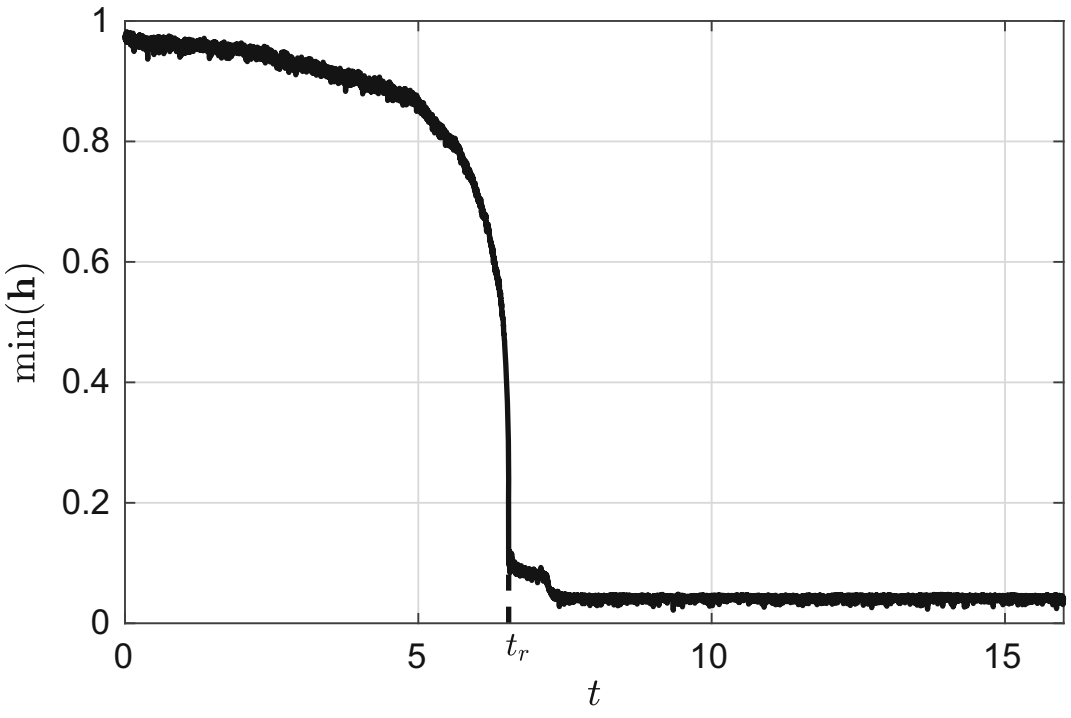

**Fig. 9** Mean rupture time $\bar{t_r}$ as a function of the noise intensity $\sqrt{2T}$. Error bars show the standard deviation of $\bar{t_r}$

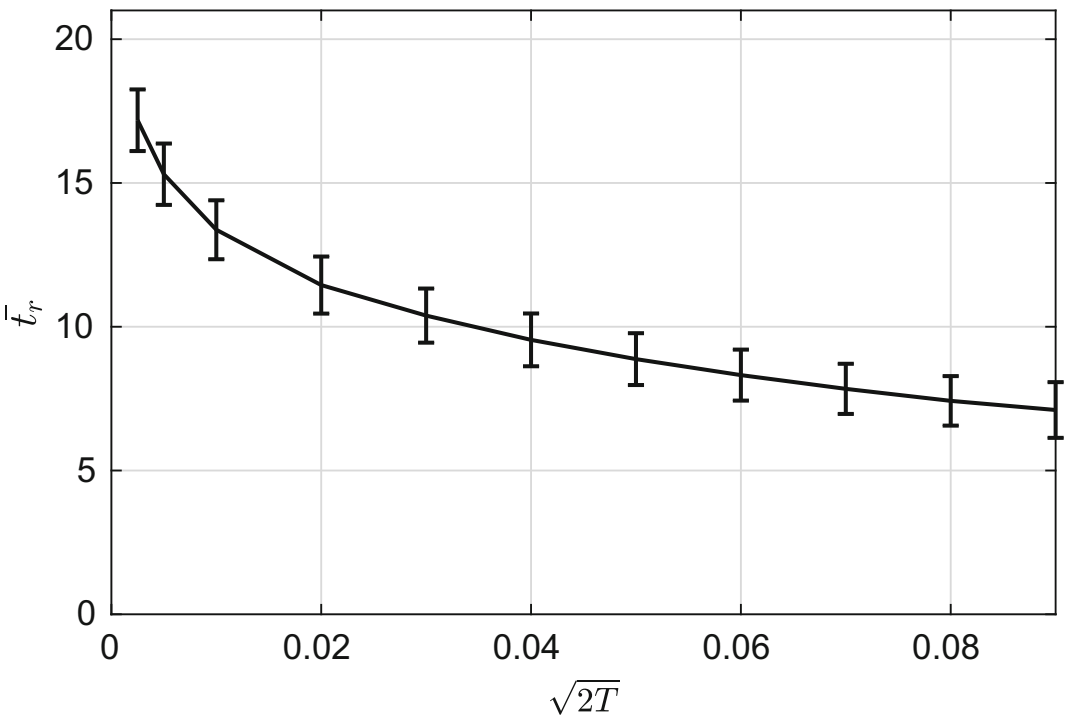

$$t_r = \left( \sup_{\forall n} |\min(\mathbf{h}_{t_n+1}) - \min(\mathbf{h}_{t_n})| \right) \Delta t_s. \tag{73}$$

The reason behind this choice lies in the fact that the minimum film height changes rapidly in the vicinity of the film rupture time as can be seen in Fig. 8. Using this definition of the film rupture time, we can then compute its dependence of the noise intensity. For each value of $T$ we set the number of sample paths. The variation of $\bar{t_r} = \mathbb{E}(t_r)$ with $\sqrt{2T}$ can be seen in Fig. 9. From the same figure it is also evident that the noise has the effect of reducing the time it takes on average for the film to breakup, as expected from physical intuition. Nevertheless, as the noise intensity increases, the rupture time tends to saturate to a fixed value. This saturation time seems to reflect the fact that the system has a proper time scale to react to fluctuations, no matter how intense they are.

## 6 Concluding Remarks

In this work we have introduced an alternative and rigorous derivation of the stochastic thin-film equation from first principles. We have also presented numerical simulations to study the stability of an initially homogeneous state and the response of the system to fluctuations. The starting point of the derivation is the stochastic Navier–Stokes equations for the velocity field of an incompresible fluid on a planar horizontal solid substrate. The relation of these equations with an underlying Hamiltonian dynamics of the constituent particles of the fluid is sketched in Fig. 1, where we also highlight our contribution to the state-of-the-art and summarise all possible model equations obtained from the original Hamiltonian system. Having the fluctuating equations governing the relevant hydrodynamic fields for the system dynamics, we apply the widely-known long-wave approximation, which makes possible a considerable reduction of the dynamics in terms of the height of the liquid film. Despite such a simplification, the resultant SPDE describing the temporal evolution of the height of the film along the streamwise direction contains a fluctuating term which is not convenient for practical purposes. Inspired by the work of Davidovitch et al. [12] and Grün et al. [37], we propose a tractable, but general, state-dependent noise term to replace the original one. At this point, we analyse the condition the noise term must fulfil in order to satisfy the detailed-balance condition. We show that, for the new equation to fulfil the detailed-balance condition the noise must have the same structure as the one proposed in previous works [12,37]. We subsequently justify the structure of the alternative noise term proposed by using a $Q$-Wiener representation of the original noise. We then show that the original stochastic dynamics and the alternative SPDE converge when the long correlation-length limit is imposed along the wall-normal direction. That is, both noise terms produce an equivalent statistics when the simpler is considered to be perfectly correlated along the wall-normal direction. We believe this limit is more physically meaningful than the uncorrelated noise originally derived by Grün et al. [37].

We also demonstrate the gradient-flow structure of the thin-film equation and define the associated energy functional, $\mathcal{H}$. By studying the variation of such a functional, we show that an initially spatially homogeneous film is unconditionally linearly stable to sufficiently small perturbations in the case of a negligible interface potential. In the case of a general interface potential, $\phi(h) = \alpha\, h^{-3} - \beta\, h^{-2}$, which is the sum of a non-negative convex and a concave term, we find that the condition required for the film to become unstable is $\beta > 2\alpha$. This result is crucial in that it gives the conditions under which the dewetting process can occur. To scrutinse the nonlinear dynamics of the film and resulting rupture process, we propose a numerical algorithm based on a spectral collocation method. We perform simulations of the dynamics of the thin film and study the evolution of the energy functional close to the rupture of the film. Finally, we study the effect of the noise intensity on the rupture time, and our results are in agreement with those by Grün et al. [37]. As a remark, we observe that the rupture time seems to saturate as the noise intensity increases, resembling the saturation of the escape time in some thermally-activated processes, e.g. nucleation.

**Acknowledgements** The work of RG is supported by an Imperial College London (ICL) President's PhD Scholarship and by ICL's Centre for Doctoral Training in Fluid Dynamics Across Scales. We acknowledge financial support from the European Research Council via Advanced Grant No. 247031 and from the Engineering and Physical Sciences Research Council through Grants No. EP/L025159, EP/L020564, EP/P031587 and EP/L024926.